\documentclass{article}

\PassOptionsToPackage{numbers, compress}{natbib}

\usepackage[final]{neurips_2025}

\usepackage[utf8]{inputenc} 
\usepackage[T1]{fontenc}    
\usepackage{hyperref}       
\usepackage{url}            
\usepackage{booktabs}       
\usepackage{amsfonts}       
\usepackage{amsmath}
\usepackage{nicefrac}       
\usepackage{microtype}      
\usepackage{xcolor}         
\usepackage{xpatch}
\usepackage{graphicx}
\usepackage{tikz}

\usepackage[title]{appendix}

\usepackage{xpatch}
\makeatletter
\xapptocmd{\NAT@bibsetnum}{\setlength{\leftmargin}{0pt}\setlength{\itemindent}{\labelwidth}\addtolength{\itemindent}{\labelsep}}{}{}
\makeatother

\title{Surrogate-Based Differentiable Pipeline for Shape Optimization}

\author{%
Andrin Rehmann \quad Nolan Black \quad Josiah Bjorgaard \quad Alessandro Angioi\\ \quad \textbf{Andrei Paleyes} \quad \textbf{Niklas Heim} \quad \textbf{Dion Häfner} \quad \textbf{Alexander Lavin}\\
Pasteur Labs\\
\texttt{\{firstname.lastname\}@simulation.science}
}

\begin{document}

\maketitle

\begin{abstract}
  Gradient-based optimization of engineering designs is limited by non-differentiable components in the typical computer-aided engineering (CAE) workflow, which calculates performance metrics from design parameters. While gradient-based methods could provide noticeable speed-ups in high-dimensional design spaces, codes for meshing, physical simulations, and other common components are not differentiable even if the math or physics underneath them is. We propose replacing non-differentiable pipeline components with surrogate models which are inherently differentiable. Using a toy example of aerodynamic shape optimization, we demonstrate an end-to-end differentiable pipeline where a 3D U-Net full-field surrogate replaces both meshing and simulation steps by training it on the mapping between the signed distance field (SDF) of the shape and the fields of interest. This approach enables gradient-based shape optimization without the need for differentiable solvers, which can be useful in situations where adjoint methods are unavailable and/or hard to implement.
\end{abstract}

\section{Introduction}
Many computer-aided engineering (CAE) problems revolve around finding an optimal geometry. Examples include structural optimization (designing lightweight yet robust structures), aerodynamic shape optimization (maximizing lift-to-drag ratio), heat exchanger design (improving heat dissipation), and acoustic optimization (reducing resonance). The underlying challenge is common across domains: given a parametrized family of shapes and a performance metric, find the design parameters that maximize (or minimize) the quantity of interest.

Traditionally, such optimization tasks rely heavily on manual iteration: an engineer proposes a design based on physical intuition, generates a mesh, runs a simulation, evaluates quantities of interest (QoIs), and iteratively modifies the \emph{design parameters}. This workflow can be summarized as a repeated application of the following pipeline:
\begin{equation*}
    \texttt{Design Parameters} \rightarrow
    \texttt{3D geometry} \rightarrow
    \texttt{Meshing} \rightarrow
    \texttt{Simulation} \rightarrow
    \texttt{QoI}
\end{equation*}

Due to the high-dimensional design spaces typical of engineering applications, optimization methods informed by intuition rather than gradients do not scale well.  Gradient-based methods promise higher efficiency but require the entire pipeline to be differentiable. Commercial and open-source design-space software like computer aided design (CAD) tools, meshing tools~\cite{geuzaine2009gmsh}, and simulation software~\cite{openfoam} are often not natively differentiable. Recovering adjoint sensitivities from solvers is often time-consuming and error-prone; meanwhile, newer projects leveraging automatic differentiation offer flexibility. In many cases, however, gradients can only be obtained for mesh quantities but not for the mesh structure itself. Gradient-based geometry optimizations have been done, but they are one-off solutions that do not generalize to new software components \cite{wintzer2015under}, or they apply to a subset of the workflow steps described above \cite{guillard2024deepmesh, xue2023jax}. It is imperative to automate the full chain of steps in order to avoid manual set-up in the design space software, which would drastically increase the step time of algorithmic optimization. 

By training surrogates on parts of the workflow, we can replace non-differentiable components, with data-driven approximations. The benefits of surrogates are threefold: (i) they are inherently differentiable; (ii) the input geometry representation can be chosen freely; and (iii) the neural networks can provide fast approximation of the results \cite{Lavin2021SimulationIT}, which are often sufficient to compute gradients for optimization \cite{guillard2024deepmesh}.

\textbf{Contributions.} In this work, we demonstrate this approach on a concrete aerodynamic shape optimization problem. While the example is simplified, the methodology is general:
\begin{itemize}
    \item We show how to construct an end-to-end differentiable optimization pipeline by replacing OpenFOAM~\cite{openfoam} with a U-Net surrogate.
    \item We provide a modular pipeline for shape optimization where each component can be independently substituted depending on the problem of interest.
    \item We validate the Tesseract framework's~\cite{tesseract-core} ability to compose pipeline components (differentiable and non-differentiable) into end-to-end workflows for data-generation, model training, and optimization.
\end{itemize}

The Tesseract framework  mentioned above provides a modular architecture for composing computational pipelines from independent components. Crucially, Tesseracts expose gradient information at the component level, making it easy to compose them via AD frameworks like JAX or PyTorch. Each of the components listed below is implemented as a Tesseract.

\section{Methodology}

In this section we give an overview of our end-to-end differentiable pipeline, and explain key design choices we made while building it. The pipeline is split into three parts: data generation, training, and optimization --- each consisting of multiple components.

\textbf{Data generation.} The data generation pipeline consists of four components. First, \texttt{geometry} generates 3D geometries, both a signed distance field (SDF) on a regular grid and a surface mesh, from design parameters. The geometries are truncated cones with hemispherical caps, arbitrarily rotated in space. Next, \texttt{mesh} converts the surface mesh into a volumetric tetrahedral mesh using GMSH~\cite{geuzaine2009gmsh}, with adaptive refinement near the shape's boundary. Then, the \texttt{openfoam} component uses OpenFOAM to solve the incompressible Navier-Stokes equations to obtain flow fields around the shape. Finally, \texttt{interpolate-to-grid} maps the volumetric mesh data into a regular grid suitable for surrogate training. Full details on geometry representation, meshing, and computational fluid dynamics (CFD) simulation are provided in the appendix.

\textbf{Surrogate training.} To replace the computationally expensive and non-differentiable CFD simulation, we trained a surrogate model to predict flow fields directly from the SDF representation of the geometry. The architecture chosen for the surrogate model is a U-Net\cite{oktay2018attention} leveraging the implementation in the PhysicsNemo library\cite{physicsnemo2023} with additional modifications. The network receives the SDF as input, as well as enriched input representations: positional encoding via sine and cosine functions of normalized spatial coordinates, and obstacle mask distinguishing interior from exterior regions (more details on input can be found in Appendix~\ref{sec:rich-input}). The \texttt{unet-trainer} component trains a U-Net to learn a direct mapping from the enriched input to the flow fields.

\textbf{Optimization.} The optimization workflow chains \texttt{geometry}, \texttt{interpolate-to-grid}, and \texttt{unet-inference} to predict flow fields from design parameters. The loss is computed, and its gradient with respect to the design parameters is then derived through backpropagation.

The modularity of this formulation means that individual components can be swapped or upgraded independently. For example, the geometry representation could be changed from rounded cones to parametric airfoils, or the U-Net could be replaced with a different architecture. This approach enables practitioners to selectively introduce differentiability where it provides the most value, rather than having to build monolithic differentiable solvers (often from scratch).

\begin{figure}
    \centering
    \includegraphics[width=\linewidth]{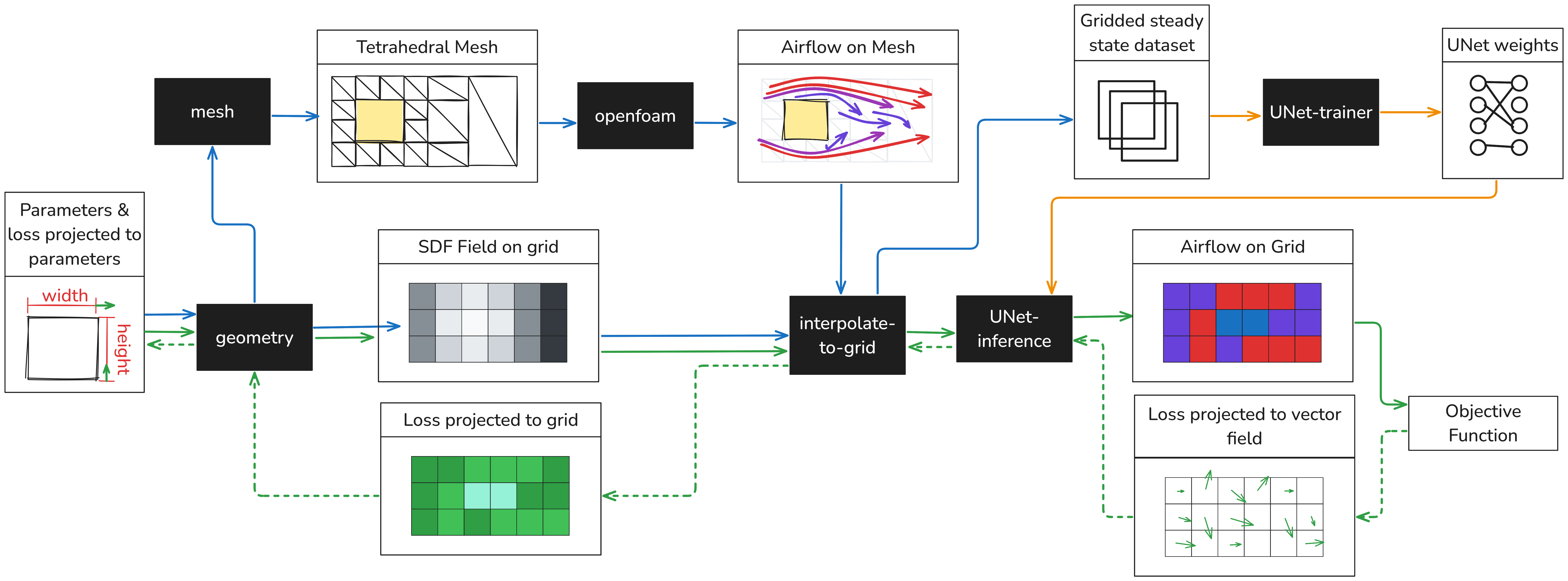}
    \caption{Pipeline architecture for surrogate-based shape optimization. Arrow colors indicate workflow stages: blue for data generation, orange for surrogate training, green for optimization. Solid arrows represent forward primal values, while dashed arrows represent backward gradient flow. All components that are written as Tesseracts are displayed as black boxes.}
    \label{fig:illustration}
\end{figure}



     

\section{Demonstration}

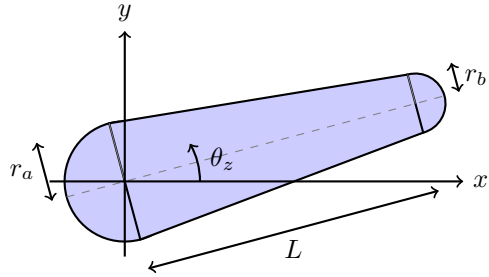
\begin{figure}[h]
\centering
\begin{tikzpicture}[scale=1.0]

\begin{scope}[rotate=15]
  \draw[thick, fill=blue!20] (0,0) circle (0.8);

  \draw[thick, fill=blue!20] (4,0) circle (0.4);

  \draw[thick, fill=blue!20] (0,0.8) -- (4,0.4) -- (4,-0.4) -- (0,-0.8) --
  cycle;

  \draw[dashed, gray] (-0.8,0) -- (4.4,0);

  \draw[<->, thick] (-1.0,0) -- (-1.0,0.8) node[midway, left] {$r_a$};
  \draw[gray] (0,0) -- (0,0.8);   

  \draw[<->, thick] (4.6,0) -- (4.6,0.4) node[midway, right] {$r_b$};
  \draw[gray] (4,0) -- (4,0.4);   

  \draw[<->, thick] (0,-1.2) -- (4,-1.2) node[midway, below] {$L$};

\end{scope}

\draw[->, thick] (-1,0) -- (4.5,0) node[anchor=west]{$x$};
\draw[->, thick] (0,-1) -- (0,2) node[anchor=south]{$y$};

\draw[->, thick] (1,0) arc (0:30:1);
\node[black] at (1.3,0.3) {$\theta_z$};
\end{tikzpicture}
\caption{Example of the primitives we use for shape optimization}
\label{fig:geometry}
\end{figure}

\begin{figure}[h]
\centering
\includegraphics[width=0.48\linewidth]{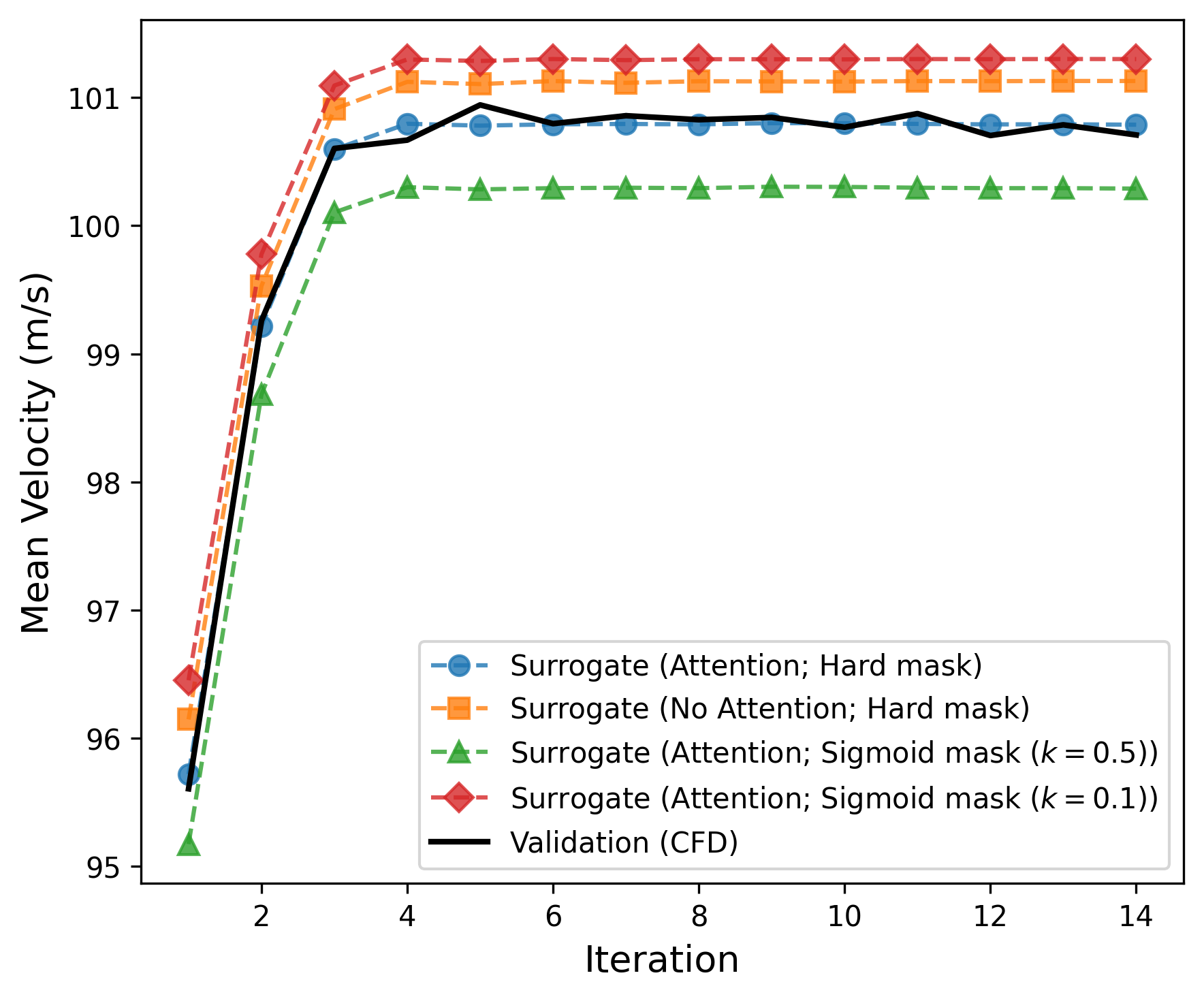}
\hfill
\includegraphics[width=0.48\linewidth]{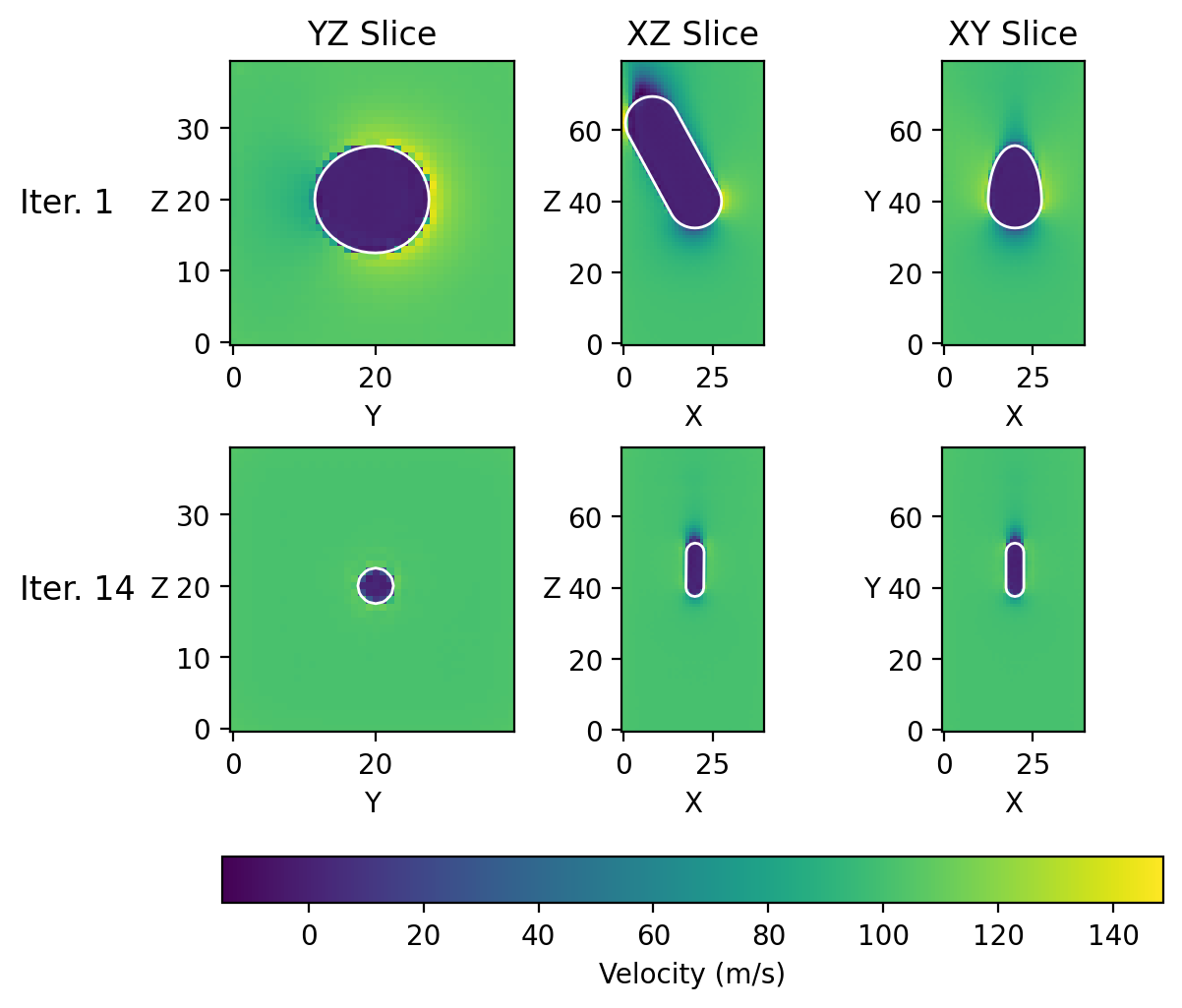}
\caption{Optimization results: (left) convergence of the objective function over 14 iterations; (right) design evolution showing the shape aligning with flow and reducing volume.}
\label{fig:optimization}
\end{figure}

To demonstrate the effectiveness of our proposed optimization pipeline, we consider the design presented in \autoref{fig:geometry}, which describes a 3D rounded cone. It is parametrized by two radii $r_a$, $r_b$, a length, $L$ and the angle $\theta_z$. The geometry and the parameter ranges used for data generation are described in more detail in \ref{sec:geometry}. We consider this shape as being immersed in a fluid flow, and the target for optimization is chosen to maximize the mean $x$-component of the flow velocity. This effectively minimizes the drag created by the shape.

We first generated a dataset using the data generation pipeline (see \autoref{fig:illustration}) to create 896 samples that were split randomly into 768 and 128 samples for test and validation. We then trained a U-Net surrogate model to predict fluid flow around the solid object. The model predicts the 3-component velocity vector field with min-max normalization to a magnitude range of [0, 128.6749]. Training was performed on gridded data with spatial resolution of $80 \times 40 \times 40$ (after slicing from full domain), which centers a box of this size on the solid object and neglects the external boundary. The target data is a 3-component velocity field. The model hyperparameters are detailed in Appendix~\ref{sec:hyperparams}. We performed ablation studies for four distinct configurations to determine the effects of attention and obstacle mask smoothing as detailed in Appendix~\ref{sec:ablation}.

The optimization workflow is made differentiable through a system of Tesseracts. It is summarized as: 
\begin{align*}
    & \texttt{design-params} \leftrightarrow
    \texttt{geometry} \leftrightarrow 
    \texttt{interpolate-to-grid} \\
    & \qquad \leftrightarrow 
    \texttt{unet-inference} \leftrightarrow 
    \texttt{QoI},
    \label{eq:optimization_workflow}
\end{align*}
where $\texttt{QoI}$ is defined as the scalar objective function $\varTheta = \mathrm{mean}(U_x)$, and double arrows indicate forward propagation of primals and backward propagation of gradients. Gradient-based optimization of $\varTheta$ is performed using the MMA Optimizer \cite{svanberg1987method}. The design parameters $\{r_a, r_b, L, \theta_z \}$ are initialized as $r_a^0 = r_b^0 = 1.5$, $L^0 =5$ and $\theta_z^0= 0.5$ and bounded according to $r_a \in (0.5, 1.5)$, $r_b \in (0.5, 1.5)$, $L \in(2.0, 5.0)$, and $\theta_z \in  (-0.5,0.5)$, then optimization was performed through the iterative maximization of $\varTheta$ for 20 iterations or until the relative change in design variables is under 1\%. 

The optimization convergence behavior and the design evolution are illustrated in \autoref{fig:optimization}. As expected, the optimization alters the design so that the solid object creates as little drag as possible to maximize the average flow velocity by minimizing the frontal area of the object. The optimized result, therefore, has aligned the inclusion with the flow and decreased its volume to the design limits. Although the design space is rather limited, the gradient recovery was successful, and we observe similar optimization trajectories for all surrogate model variants; convergence was nearly monotonic and reached the desired tolerance in less than 14 iterations. The demonstrated objective has a straightforward solution, and future work will explore more complex geometries, different objectives, and generalization performance outside of the training regime. This study demonstrates that the Tesseract workflow enabled the recovery of gradients from a previously non-differentiable workflow, and those gradients were sufficient to explore the parameterized design space.

\section{Conclusions}
We demonstrated a surrogate-based approach to gradient-guided shape optimization that does not require a differentiable physics solver by wrapping computational units into Tesseracts. While the specific problem we investigated uses a limited design space, the modularity of this pipeline makes it extensible with relative ease to more interesting problems, especially ones with more complex geometries and loss functions. It is worth noticing, however, that this strategy requires substantial upfront investment in data generation and training, and introduces model risk from approximation errors that necessitate some form of validation of optimized designs with high-fidelity simulations.

\bibliographystyle{unsrtnat}
\small
\bibliography{Reference}

\newpage
\begin{appendices}
\section{Geometry representation and parametrization}

\subsection{\texttt{geometry} Tesseract} 
\label{sec:geometry}

In order to generate a family of candidate shapes, we utilized a differentiable signed distance function (SDF) representation. For the sake of definiteness, we use truncated cones with hemispherical caps as the primitive for both data generation and optimization.

The primitives are parameterized by six design variables, and we generated 1000 geometries with the following ranges:

\begin{enumerate}
    \item Two radii ($r_a$, $r_b$) in radians controlling both the cone's cross-sectional dimensions and the curvature of the hemispherical caps at the base and tip. Uniformly sampled from the range [0.5, 1.5].
    \item A scalar length ($L$) defining the cone's extent. Uniformly sampled from the range [2.0, 5.0].
    \item Three Euler angles ($\theta_x$, $\theta_y$, $\theta_z$) in radians specifying the cone's spatial orientation. The cone's axis is initially aligned with the x-axis and then rotated according to the specified Euler angles. Here, only the angle $\theta_y$ is sampled from the range [-0,5 0.5], where the $x$ and $y$ components are set to zero.
\end{enumerate}

The geometries are exported as both SDFs on a regular grid and surface meshes; this is because the former are more amenable to automatic differentiation, whereas the latter are needed by the physical solver.

We evaluate the SDFs for each geometry on a regular grid in a vectorized fashion using JAX. The grid spans a bounding box specified by user-defined bounds. Notice that for this part of the computation, automatic differentiation is exposed through the vector-Jacobian product interface of Tesseracts.

Surface meshes are extracted from the SDF field using the marching cubes algorithm with the Lewiner~\cite{lewiner2012} variant, which ensures topological correctness. Finally, the mesh is post-processed using Laplacian smoothing to reduce staircase artifacts from the discrete grid representation while preserving the overall geometry.

\subsection{\texttt{mesh} Tesseract}
The volumetric mesh generation component converts the parameterized surface geometry into a domain suitable for CFD analysis. This process uses GMSH to create a tetrahedral mesh representing the fluid domain around the obstacle.

The computational domain consists of a rectangular box with the optimized shape positioned as an interior void. The box is centered at $(L_x/4, 0, 0)$, where $L_x$ is the projected length of the shape on the $x$ axis.

The domain is constructed using GMSH's built-in geometry kernel. To ensure high-quality volume mesh generation, the input STL surface undergoes optional reparameterization. This process uses GMSH's \texttt{classifySurfaces} function with a tolerance angle ($20^\circ$) to identify distinct geometric features and create smooth surface patches. Curves are split when they deflect beyond a specified angle threshold ($180^\circ$).

Mesh refinement is controlled through a hierarchy of size parameters:
\begin{itemize}
  \item Global bounds constrain element sizes throughout the domain
  \item A coarse target size is applied to far-field regions
  \item A fine target size is enforced on the obstacle surface
\end{itemize}

This approach concentrates the mesh's resolution near the shape boundary where more detail is needed, while maintaining efficiency in the far field.

The resulting tetrahedral mesh is exported in GMSH MSH 2.2 format. Physical groups are assigned to all boundary surfaces and the  fluid volume to facilitate boundary condition specification in the CFD solver.

\section{Computational Fluid Dynamics Simulation}

The CFD simulation component uses OpenFOAM to solve the incompressible Navier-Stokes equations. This module accepts the tetrahedral mesh from the meshing stage and returns flow field data for subsequent computation.

The simulation workflow is orchestrated through a bash script that executes the standard OpenFOAM preprocessing, solving, and postprocessing pipeline. The GMSH mesh is first converted to OpenFOAM's native format using the \texttt{gmshToFoam} utility, which preserves the physical group tags assigned during mesh generation.

We used the \texttt{incompressibleFluid} solver within OpenFOAM's \texttt{foamRun} framework to solve the incompressible Navier-Stokes equations. The flow is characterized by a density $\rho = 1 \,\text{kg}/\text{m}^3$ and kinematic viscosity $\nu = 10^{-5} \text{m}^2/\text{s}$. In this regime, the  Reynold's number remains on the order of $10^7$ regardless of the inclusion geometry, so Reynolds-averaged Navier–Stokes (RANS) equations are used to describe the fluid flow.

The computational domain employs the following boundary conditions:
\begin{itemize}
  \item Freestream velocity conditions set to the uniform flow velocity $(100, 0, 0) \, \text{m}/\text{s}$ in both the inlet and the outlet.
  \item No-slip conditions enforcing zero velocity at the channel walls
  \item No-slip condition representing the solid boundary at the obstacle surface.
\end{itemize}

The solver employs an adaptive time-stepping scheme with a maximum Courant number of 5 and an initial time step of $\Delta t = 0.001$ s. The simulations progress from $t=0$ to $t=0.15$ s, with the output written at the initial and final times. This transient formulation allows the flow to develop from the uniform initial conditions to a quasi-steady state suitable for force evaluation.

Upon completion, the simulation data is exported to VTK format using OpenFOAM's \texttt{foamToVTK} utility. The output includes separate files for each boundary patch (inlet, outlet, walls, obstacle surface) and the internal volume mesh, each containing point coordinates, cell connectivity, and field data (velocity, pressure, turbulence quantities). These VTK files are read using PyVista~\cite{sullivan2019pyvista} and converted to a structured format containing both geometric information and field quantities, enabling subsequent interpolation and gradient computation in the optimization pipeline. Since some simulation runs failed or resulted in poor convergence. Therefore, we removed samples with NaN values or with mesh cells that had a velocity vector with a magnitude greater than 160.

\section{Surrogate Training}

\subsection{Input engineering}
\label{sec:rich-input}

Rather than using only the SDF, the network receives an enriched input representation:
\begin{itemize}
  \item SDF: Normalized signed distance values indicating proximity to the obstacle
  surface.
  \item Positional encoding: Sine and cosine functions of normalized spatial coordinates
  $(x, y, z)$, providing explicit positional information to the translation-invariant convolutions.
  \item Obstacle mask: Indicator distinguishing interior (SDF < 0) from exterior regions
   - Hard mask: Binary threshold at SDF=0
   - Sigmoid mask $m$ with temperature parameter $k$
     where smaller $k$ produces sharper transitions at the boundary
The sigmoid mask provides a differentiable alternative to hard thresholding, enabling gradients to flow through the masking operation during backpropagation.
     
\end{itemize}     
\begin{equation}
       m(\text{SDF}) = \sigma\left(-\frac{\text{SDF}}{k}\right) = \frac{1}{1 + e^{\text{SDF}/k}}
\end{equation}

\subsection{Hyperparameters}
\label{sec:hyperparams}

Key hyperparameters for model and optimizer were chosen through trial and error to establish stable baseline training.
\begin{itemize}
    \item Batch size: 64
    \item Learning rate: $1.5*10^{-4}$
    \item Optimizer: Adam with default $\beta$ parameters ($\beta_1$=0.9, $\beta_2$=0.999)
    \item Training epochs: 400
    \item Loss function: Mean squared error (MSE) on predicted velocity fields
    \item Encoder depth: 3 levels with progressively increasing feature maps [64, 64, 128, 128, 512, 512]
    \item Attention Gates: Feature maps [512, 128] with 256 intermediate channels 
    \item Convolutional blocks: 2 consecutive blocks per level using 3×3×3 kernels with GELU activation
    \item Normalization: Layer Normalization
    \item Pooling: 3D max pooling with 2×2×2 stride between encoder levels
\end{itemize}

\begin{figure}
    \centering
    \includegraphics[width=1.0\linewidth]{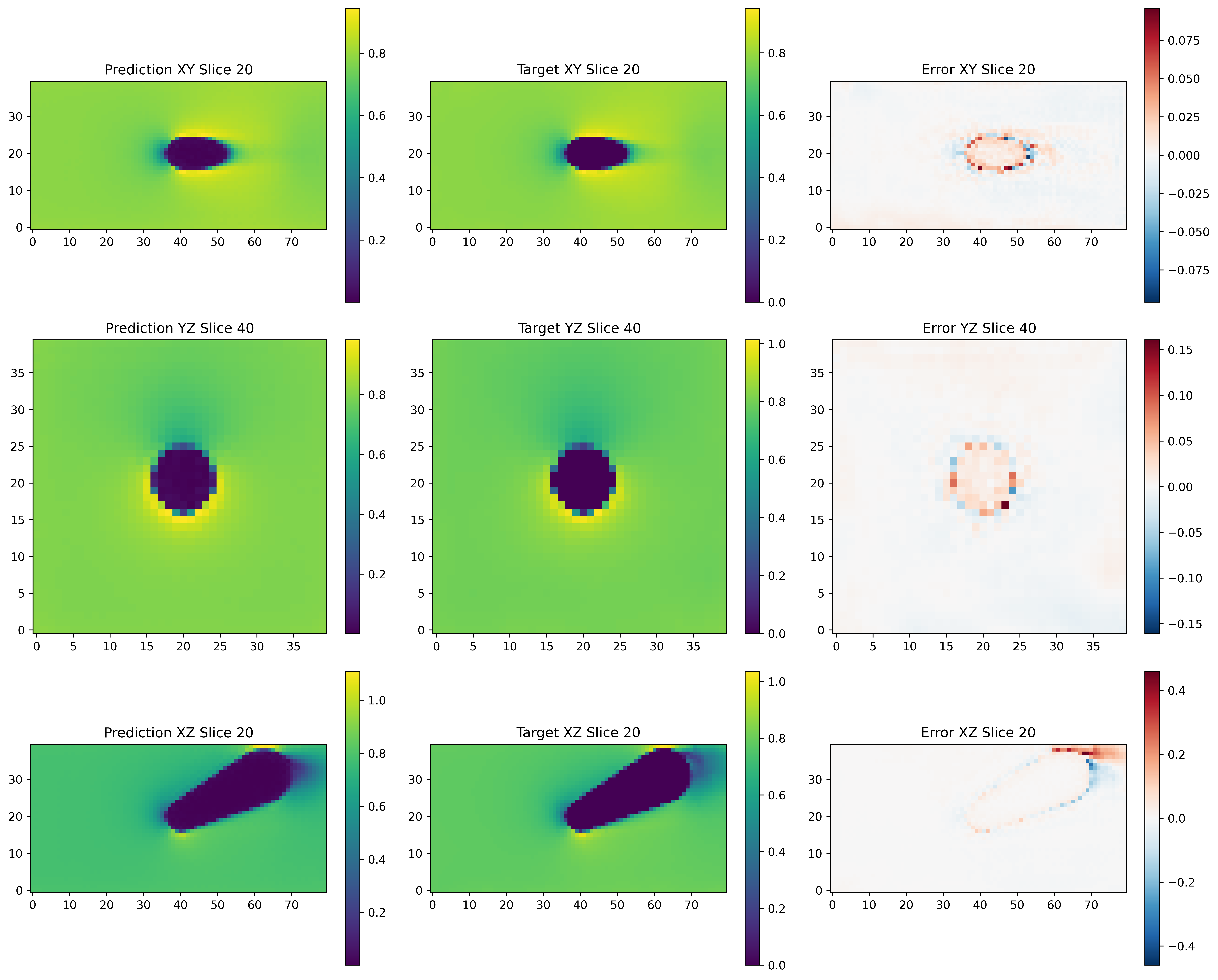}
    \caption{Example normalized target and prediction slices in xy, yz, and xz planes for trained baseline model.}
    \label{fig:ml_slices}
\end{figure}

Training was performed on 4 NVIDIA A100 80GB PCIe GPUs with an approximate training time of 6.5 hours per run, and results are illustrated in \autoref{fig:ml_slices}.

\subsection{Ablation Studies}
\label{sec:ablation}

\begin{figure}
    \centering
    \includegraphics[width=\linewidth]{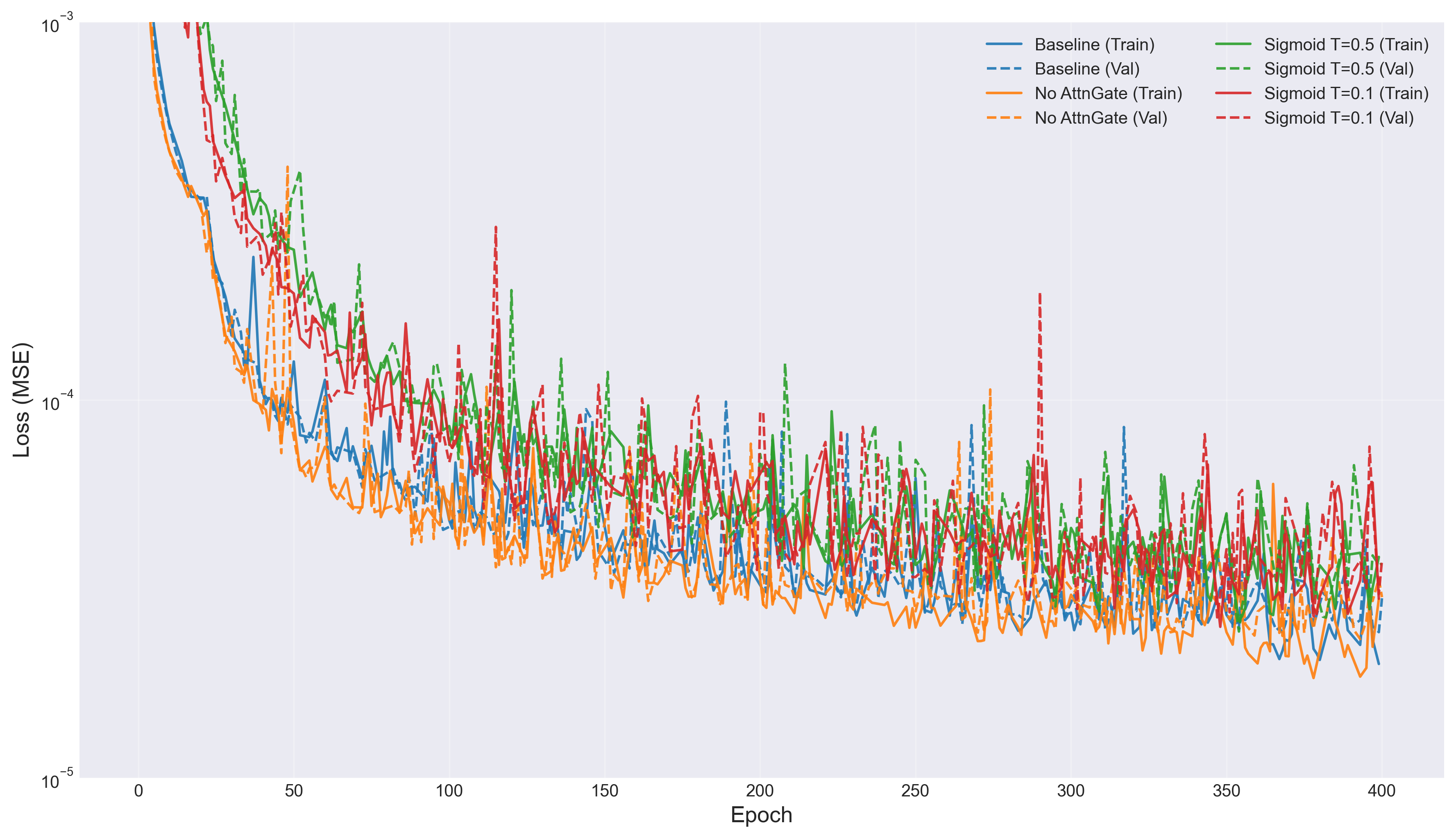}
    \caption{Training and validation loss curves for the ablation study. Solid lines indicate training loss, dashed lines indicate validation loss. All models converge to similar performance, with the baseline configuration achieving the lowest validation loss.}
    \label{fig:train_val}
\end{figure}

\begin{figure}
    \centering
    \includegraphics[width=1.0\linewidth]{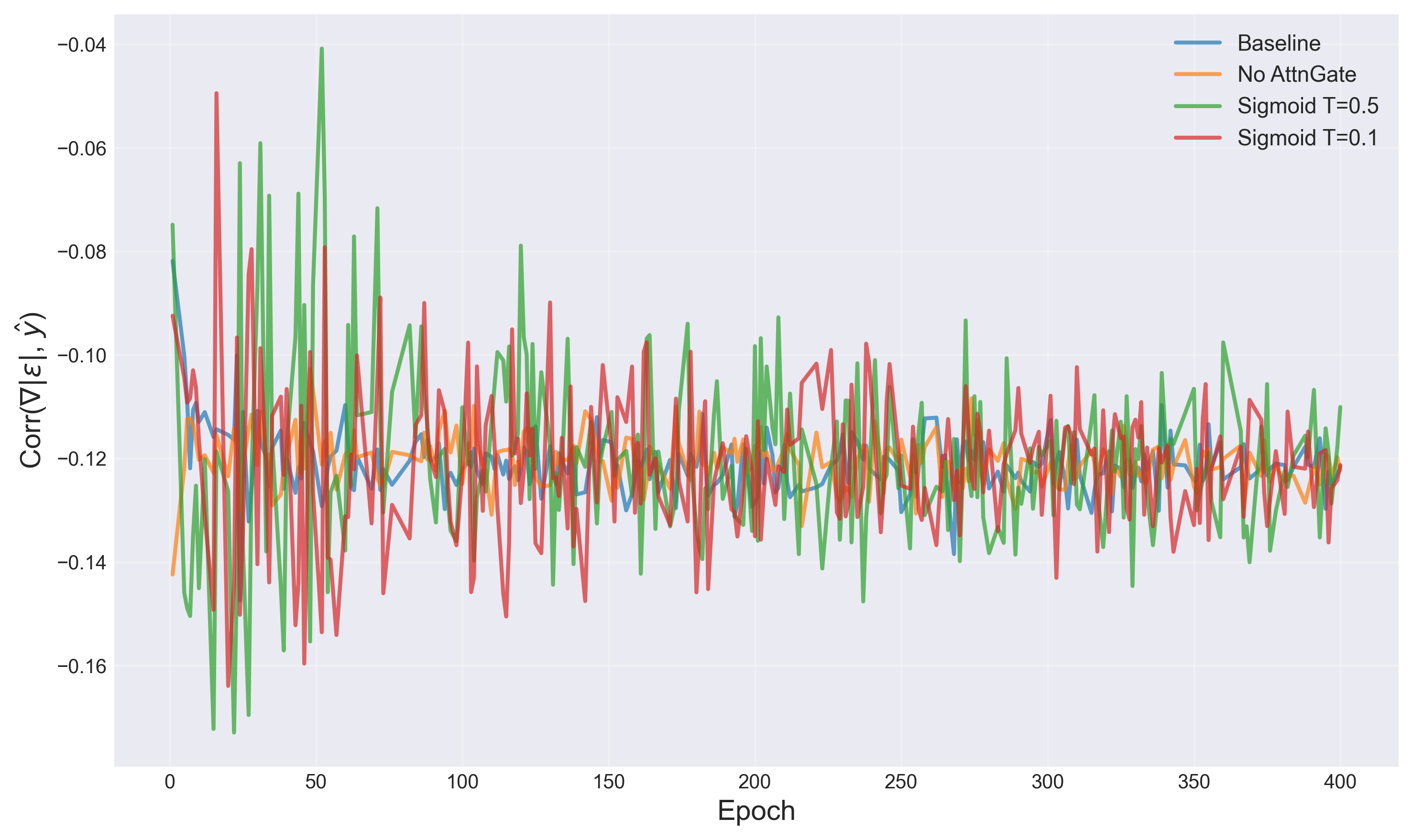}
    \caption{Error gradient correlation metric during training. The metric measures the Pearson correlation between spatial gradients of the absolute error field ($\nabla\epsilon$) and the predicted velocity values ($f(x)$). Negative values indicate that regions with rapidly varying errors tend to have lower predicted velocities, suggesting difficulty in low-velocity or boundary regions.}

    \label{fig:train_gradient}
\end{figure}

We performed experiments for four distinct configurations to determine the effects of attention and obstacle mask smoothing as detailed in \autoref{tab:training_results}. In addition to MSE loss (\autoref{fig:train_val}), we calculated the correlation between the spatial gradient of the absolute error $\epsilon=|y-f(x)|$ and the predicted value $f(x)$:
\begin{equation}
    \text{Corr}(\nabla \epsilon, f(x))
\end{equation}
where $Corr$ is the Pearson correlation function, $\nabla$ is the gradient operator (calculated by nearest neighbor finite difference). This metric assesses whether regions of fluctuating error align with regions of high velocity (\autoref{fig:train_gradient}). 

These results lead to the following conclusions.
\begin{itemize}
    \item Attention did not lead to improved model performance in this particular dataset.
    \item Sigmoid masking resulted in slower model convergence, regardless of $k$, but did not strongly affect the overall performance of the trained model.
    \item Stronger fluctuations in model error had a slightly higher likelihood to be found in regions of low velocity than in regions of high velocity.
\end{itemize}

\begin{table}[h]
\centering
\caption{Training Results: Ablation Study Comparing Attention Gates and Masking Strategies}
\label{tab:training_results}
\begin{tabular}{lcccccc}
\toprule
Attn & Mask Type & Temp & Train Loss & Val Loss & Best Val & Corr($\nabla|\epsilon|$,$\hat{y}$) \\
& & & (Final) [$\times 10^{-5}$] & (Final) [$\times 10^{-5}$] & (Train/Val, Epoch) [$\times 10^{-5}$] & \\
\midrule
Yes & Hard & \texttimes & 2.00 & 2.99 & 2.17/2.42 (368) & -0.1217 \\
No & Hard & \texttimes & 2.97 & 3.01 & 1.85/2.33 (393) & -0.1223 \\
Yes & Sigmoid & 0.5 & 3.84 & 3.69 & 2.61/2.44 (354) & -0.1101 \\
Yes & Sigmoid & 0.1 & 2.98 & 3.77 & 2.82/2.63 (378) & -0.1213 \\
\bottomrule
\end{tabular}
\end{table}

\end{appendices}
\end{document}